\documentclass[%
 reprint,
 superscriptaddress,
 amsmath,amssymb,
 aps,
 pra,
]{revtex4-2}

\usepackage{color}
\usepackage{mathrsfs}
\usepackage{graphicx}
\usepackage{dcolumn}
\usepackage{bm}
\usepackage{hyperref}

\begin{document}

\preprint{APS/123-QED}

\title{Radiative polarization dynamics of relativistic electrons in an intense electromagnetic field}

\author{Yuhui Tang}
 \affiliation{State Key Laboratory of Nuclear Physics and Technology, and Key Laboratory of HEDP of the Ministry of Education, CAPT, School of Physics, Peking University, Beijing 100871, China}
\author{Zheng Gong}
 \affiliation{State Key Laboratory of Nuclear Physics and Technology, and Key Laboratory of HEDP of the Ministry of Education, CAPT, School of Physics, Peking University, Beijing 100871, China}
\author{Jinqing Yu}
 \affiliation{School of Physics and Electronics, Hunan University, Changsha, 410082, China}
\author{Yinren Shou}
 \affiliation{State Key Laboratory of Nuclear Physics and Technology, and Key Laboratory of HEDP of the Ministry of Education, CAPT, School of Physics, Peking University, Beijing 100871, China}
\author{Xueqing Yan}
 \email[]{x.yan@pku.edu.cn}
 \affiliation{State Key Laboratory of Nuclear Physics and Technology, and Key Laboratory of HEDP of the Ministry of Education, CAPT, School of Physics, Peking University, Beijing 100871, China}
 \affiliation{Beijing Laser Acceleration Innovation Center, Huairou, Beijing, 101400, China}
 \affiliation{CICEO, Shanxi University, Taiyuan, Shanxi 030006, China}
 \affiliation{Institute of Guangdong Laser Plasma Technology, Baiyun, Guangzhou, 510540, China}
\date{\today}

\begin{abstract}
We propose a self-consistent model which utilizes the polarization vector to theoretically describe the evolution of spin polarization of relativistic electrons in an intense electromagnetic field.
The variation of radiative polarization due to instantaneous no photon emission is introduced into our model, which extends the applicability of the polarization vector model derived from the nonlinear Compton scattering under local constant crossed-field approximation to the complex electromagnetic environment in laser plasma interaction.
According to this model, we develop a Monte Carlo method to simulate the electron spin under the influence of radiation and precession simultaneously. Our model is consistent with the quantum physical picture that spin can only be described by a probability distribution before measurement, and it contains the entire information on the spin. The correctness of our model is confirmed by the successful reproduction of the Sokolov-Ternov effect and the comparison of the simulation results with other models in the literature. The results show the superiority in accuracy, applicability, and computational efficiency of our model, and we believe that our model is a better choice to deal with the electron spin in particle-in-cell simulation for laser plasma interaction.
\end{abstract}

\maketitle


\section{Introduction}
The state-of-the-art petawatt laser facilities~\cite{mourou2006optics,danson2015petawatt,danson2019petawatt} enable the peak intensity of a short laser pulse to reach $10^{23} \mathrm{W/cm^2}$, which makes it possible to experimentally explore the exotic quantum electrodynamics phenomena~\cite{marklund2006nonlinear,bulanov2011design,di2012extremely,zhang2020relativistic,blackburn2020radiation}. On the other hand, today's particle accelerators, like some electron synchrotrons~\cite{myers1990design,chao2008physics} and laser-driven plasma accelerators~\cite{leemans2014multi,gonsalves2019petawatt}, can easily accelerate electrons up to several $\mathrm{GeV}$. When such an intense laser beam collides with such an energetic electron beam, the quantum regime $\chi_e \gtrsim 1$ can be approached, where $\chi_e \sim \gamma E/E_\mathrm{crit}$ is the invariant quantum parameter, $\gamma$ is the Lorentz factor of the electron, $E$ is the electric field of the laser, and $E_{\mathrm{crit}} \approx 1.32 \times 10^{18} \mathrm{V/m}$ is the Schwinger limit field~\cite{schwinger1951gauge}. 
Under this condition, an electron can be substantially spin-polarized on femtosecond timescales merely after emitting a few photons~\cite{del2017spin,del2018electron,li2019ultrarelativistic,li2019electron}.
On the contrary, an electron needs to take a long time, around several hours, to obtain the considerable spin polarization in the classical regime $\chi_e \ll 1$~\cite{sokolov1986radiation}. For instance, taking advantage of the Sokolov–Ternov effect~\cite{bordovitsyn1999synchrotron}, an electron beam with a high degree of polarization can be obtained after circulating several hours in a storage ring~\cite{bauier1972radiative,camerini1975measurement,learned1975polarization,serednyakov1976study}. Spin-polarized electrons are extensively utilized in high-energy physics, for instance, probing the structure of nuclei~\cite{abe1995precision,alexakhin2007deuteron}, generating polarized photons and positrons~\cite{olsen1959photon,martin2012polarization,abbott2016production}, investigating parity nonconservation effects~\cite{prescott1978parity,labzowsky2001parity,anthony2004observation}, and studying new physics beyond the standard model~\cite{moortgat2008polarized,vauth2016beam}. 

For the ultraintense laser matter interaction, the current research for the radiative polarization of electron is still focused on theories and simulations. To simulate the radiative polarization under this condition, various numerical calculation models have been developed. One of the commonly used models is based on the Monte Carlo method with the help of an artificially selected axis, and the axis is called the spin quantization axis (SQA) in some articles~\cite{li2019ultrarelativistic,li2020polarized,geng2020spin,xue2020generation}. Specifically, the electron emits photons following the same method as the normal Monte Carlo model of photon emission where the spin polarization is not considered~\cite{kirk2009pair,duclous2010monte,ridgers2014modelling,neitz2014electron,gonoskov2015extended,lobet2016modeling,gong2019radiation,di2019improved}, and once the electron emits a photon, the spin falls on a certain state of the SQA (i.e., the state oriented in the same or in the opposite direction with respect to the direction of the SQA) determined by a random number according to the probability. This operation is equivalent to applying a spin measurement operator along the SQA on the electron immediately after the photon being emitted, which would lose all the spin information perpendicular to the SQA once a photon is emitted, since measurements of spin along different axes are incompatible. In other words, the measurement may affect and change the state of the electron spin, whereas there is no measurement operator after every time when a photon is emitted in reality. Therefore, the simulation result will be different from the reality if the SQA is not selected appropriately, and different results will be obtained from different selections of the SQA for the same simulation. This implies that the SQA model cannot contain the complete three-dimensional (3D) information on the spin polarization throughout the simulation. For this reason, the SQA model is also difficult to apply simultaneously with other theories based on the spin vector, e.g. the spin precession of the classical theory~\cite{del2018electron}.

To solve the problems in the original SQA model, an improved SQA model has also been implemented recently~\cite{li2020production}. In this improved model, the SQA is calculated according to the mean spin vector after the photon emission rather than selected artificially. In each time step, not only the emitting electron falls on a certain spin state of the SQA but also the electron that does not emit photon in this time step falls on a spin state of another calculated SQA. In this way, the improved SQA model can be applied to simulate the 3D situations. To distinguish from the original SQA model, we refer to this improved model as the 3D-SQA model in the following text.

According to the quantum electrodynamics, the spin polarization of a free electron is completely defined by three parameters, the components of the mean spin vector in the rest frame~\cite{berestetskii1982quantum}. The normalized mean spin vector is also called the polarization vector $\boldsymbol{\Xi}$. In a fully polarized pure state, $|\boldsymbol{\Xi}| = 1$; in a partially polarized mixed state, $0 < |\boldsymbol{\Xi}| < 1$; and in completely unpolarized state, $|\boldsymbol{\Xi}| = 0$. The polarization vector completely describes the spin state of the electron and corresponds to the spin vector in the classical theory. So if we utilize the polarization vector to describe the evolution of the relativistic electron spin, the complete information on spin will be kept throughout the simulation, and some models have been developed based on this idea~\cite{seipt2018theory,geng2019generalizing,guo2020stochasticity}.

In this article, we propose a self-consistent polarization vector model to theoretically describe the evolution of spin polarization of relativistic electrons in an intense electromagnetic field. 
The model is based on the quantum theory of nonlinear Compton scattering under the local constant crossed-field approximation and extends the applicability to the complex electromagnetic environment by including the variation of radiative polarization due to instantaneous no photon emission. 
According to this model, we develop a Monte Carlo method to simulate the electron spin under the influence of radiation and precession simultaneously by extending the normal Monte Carlo quantum electrodynamics radiation model~\cite{kirk2009pair,duclous2010monte,ridgers2014modelling,neitz2014electron,gonoskov2015extended,lobet2016modeling,gong2019radiation,di2019improved}.
Our Monte Carlo polarization vector (MCPV) model maintains the complete information on the spin polarization of the electron at any simulation moment, and it has a computational efficiency higher than that of the 3D-SQA model and a time step condition more relaxed than that of the 3D-SQA model. 
Compared with other continuous models described by the mean spin vector~\cite{geng2019generalizing,guo2020stochasticity}, the MCPV model has the advantage of handling photon emission and electron spin simultaneously, and is more suitable for the interaction in the quantum regime.
The correctness of the MCPV model is confirmed by the successful reproduction of the Sokolov-Ternov effect and the comparison of simulation results with other models in the literature.
For the above reasons, the MCPV model is a superior choice to be integrated into the particle-in-cell (PIC) simulation to deal with the electron spin in the laser plasma interaction.
In Sec.~\ref{sec:theories}, the basic theories of our polarization vector model are introduced. In Sec.~\ref{sec:simulation-method}, the simulation method of our model is described in detail. In Sec.~\ref{sec:results-discussions}, several simulations are run to show the correctness, accuracy, and efficiency of our MCPV model. Finally, in Sec.~\ref{sec:conclusion}, conclusions are drawn.

\section{\label{sec:theories}Theories}
\subsection{Nonlinear Compton scattering}
\begin{figure*}
\includegraphics[keepaspectratio=true,width=179mm]{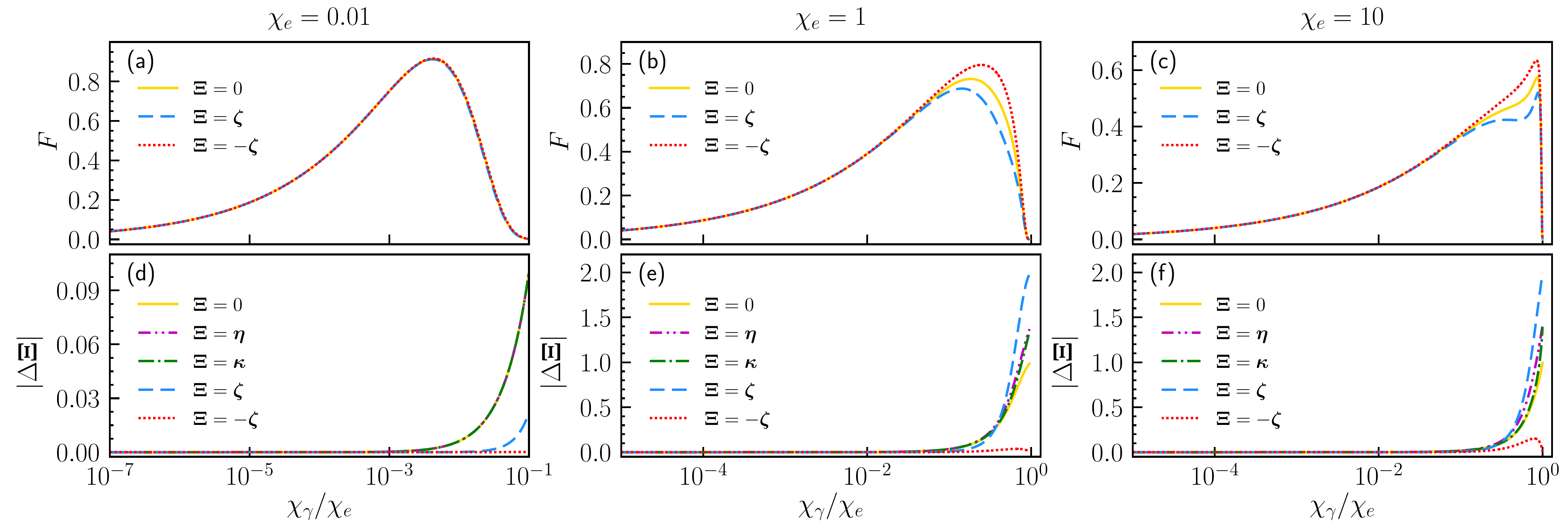}
\caption{\label{fig:delta_xi} The quantum synchrotron emissivity $F$ and the polarization variation $| \Delta \Xi |$ after the emission of a photon for [panels (a) and (d)] $\chi_e = 0.01$, [panels (b) and (e)] $\chi_e = 1$, and [panels (c) and (f)] $\chi_e = 10$. The yellow solid lines represent the initially completely unpolarized state, while the purple dash-dot-dotted lines, the green dash-dotted lines, the blue dashed lines and the red dotted lines represent the initially completely polarized states oriented in the direction of $\boldsymbol{\eta}$, $\boldsymbol{\kappa}$, $\boldsymbol{\zeta}$, and $-\boldsymbol{\zeta}$, respectively.}
\end{figure*}
In the relativistic theory, since the spin vector is not conserved in an arbitrary frame of reference, it is only possible to describe a spin state in the rest frame of the particle~\cite{berestetskii1982quantum}. Therefore, a special canonical basis $(\boldsymbol{\kappa}, \boldsymbol{\eta}, \boldsymbol{\zeta})$ defined in the rest frame is chosen for the simplest explicit analytical expressions, where $\boldsymbol{\eta} = \mathbf{E}_{\mathrm{RF}}/|\mathbf{E}_{\mathrm{RF}}|$ and $\boldsymbol{\zeta} = \mathbf{B}_{\mathrm{RF}}/|\mathbf{B}_{\mathrm{RF}}|$ are the directions of the electric and magnetic fields in the rest frame, respectively, and $\boldsymbol{\kappa} = \boldsymbol{\eta} \times \boldsymbol{\zeta}$. The electromagnetic field in the rest frame can be calculated through a Lorentz transformation from the laboratory frame, i.e.,
\begin{equation}
\begin{split}
    \mathbf{E}_{\mathrm{RF}} &= \gamma \left[ \mathbf{E} + \boldsymbol{\beta}\times\mathbf{B} - \frac{\gamma}{\gamma+1}\left(\boldsymbol{\beta} \cdot \mathbf{E} \right) \boldsymbol{\beta} \right], \\
    \mathbf{B}_{\mathrm{RF}} &= \gamma \left[ \mathbf{B} - \boldsymbol{\beta}\times\mathbf{E} - \frac{\gamma}{\gamma+1}\left(\boldsymbol{\beta} \cdot \mathbf{B} \right) \boldsymbol{\beta} \right],
\end{split}
\end{equation}
where $\boldsymbol{\beta} = \mathbf{v}/c$ is the velocity of the particle, $\gamma = 1/\sqrt{1-\boldsymbol{\beta}^2}$ is the Lorentz factor, and $\mathbf{E}$ and $\mathbf{B}$ are the electric and magnetic fields, respectively. For $\gamma \gg 1$, $\mathbf{E}_{\mathrm{RF}}$ and $\mathbf{B}_{\mathrm{RF}}$ are always orthogonal, and $\boldsymbol{\kappa}$ is oriented in the opposite direction of $\boldsymbol{\beta}$.

In a state of partial polarization, there is no definite spin wave function, but only a polarization density matrix~\cite{berestetskii1982quantum,baier1998electromagnetic,blum2012density}. The polarization density matrix $\rho$ is a $2\times 2$ matrix in the rest frame of the electron~\cite{balashov2000polarization}, and it has the following relationship with the polarization vector
\begin{equation}
\rho = \frac{1}{2}(1 + \boldsymbol{\sigma}\cdot\boldsymbol{\Xi}),\quad \boldsymbol{\Xi} = \frac{\mathrm{tr}[\boldsymbol{\sigma}\rho]}{\mathrm{tr}\rho},
\end{equation}
where $\boldsymbol{\sigma}$ is the Pauli matrix.

Two kinds of Compton scattering processes can be distinguished according to the classical nonlinearity parameter $a_0 = |e| E/\omega m_e c$, where $e$ is the electron charge, $m_e$ is the rest mass of the electron, $c$ is the speed of light, and $\omega$ is the angular frequency of the laser. For $a_0 \ll 1$ an electron interacts with only one photon (conventional Compton scattering), while for $a_0 \gg 1$ an electron interacts with a collective electromagnetic field, i.e., interacts with multiple laser photons to emit a high-energy photon (nonlinear Compton scattering)~\cite{bolshedvorsky2000polarization}.
The influence of the nonlinear Compton scattering process on the polarization density matrix of the scattered particles $\rho_f$ can be reduced to the operator $\widehat S$, and then $\rho_f$ is related to the density matrix of the initial state $\rho_i$ by the relation
\begin{equation}
    \rho_f = {\widehat S} \rho_i {\widehat S}^\dagger.
\end{equation}

For the electromagnetic wave with $a_0 \gg 1$, the formation interval of the emitted photon is so short ($\propto 1/a_0$) that the spatiotemporal variation of the background field in the region of the formation of the process can be neglected~\cite{ritus1985quantum,kirk2009pair,khokonov2010length}. Therefore, we can apply the well-known local constant crossed-field approximation and obtain the final polarization density matrix as~\cite{seipt2018theory}
\begin{equation}
    \rho = \int \rho(t,\chi_\gamma) \mathrm{d}t \mathrm{d}\chi_\gamma =  \int \frac{\alpha \mathrm{d}t \mathrm{d}\chi_\gamma}{2 \sqrt{3} \pi\tau_c \gamma \chi_e} \left(
        \begin{array}{cc}
            \tilde{\mathcal{V}}_{+} & \tilde{\mathcal{T}}^{*}\\
            \tilde{\mathcal{T}} & \tilde{\mathcal{V}}_{-}\\
        \end{array}
\right),
\end{equation}
where $\tau_c = \hbar/m_e c^2$ is the Compton time, $\alpha= e^2/\hbar c \approx 1/137$ is the fine structure constant and $\hbar$ is the reduced Planck constant. Here
\begin{eqnarray}
    &\tilde{\mathcal{V}}_{\pm} = & (2 + 3\chi_\gamma y \pm 2\Xi_{\zeta}) K_{2/3}(2y) - (1 \pm \Xi_{\zeta}) \mathrm{Int}K(2y) \nonumber \\
    & &- (\frac{\chi_\gamma}{\chi_e}\Xi_\zeta \pm 3\chi_e y) K_{1/3}(2y), \nonumber \\
    &\tilde{\mathcal{T}} = & \Xi_\kappa \left[(2+3\chi_\gamma y)K_{2/3}(2y) - (1+3\chi_\gamma y )\mathrm{Int}K(2y) \right] \nonumber \\
    & & + i \Xi_\eta \left[2 K_{2/3}(2y) - \mathrm{Int}K(2y) \right], \nonumber
\end{eqnarray}
where $K_{\nu}(x)$ is the modified Bessel functions of the second kind, $\mathrm{Int}K(x) = \int_{x}^\infty K_{1/3}(z)\mathrm{d}z$, and $y = \chi_\gamma / [3 \chi_e (\chi_e - \chi_\gamma)]$.
$\chi_e$ and $\chi_\gamma$ are the dimensionless quantum parameters of the electron and the photon, respectively, which can be calculated by
\begin{equation}
\begin{split}
\chi_e &= \frac{\gamma}{E_{\mathrm{crit}}}\sqrt{(\mathbf{E}_\perp+\boldsymbol{\beta}\times \mathbf{B})^2 + (\mathbf{E}_\parallel/\gamma)^2}, \\
\chi_\gamma &= \frac{\hbar\omega_\gamma}{m_e c^2 E_{\mathrm{crit}}}\left |\mathbf{E}_\perp+ \mathbf{\hat k}\times \mathbf{B} \right |,
\end{split}
\end{equation}
where $E_{\mathrm{crit}} = m_e^2 c^3/|e| \hbar$ is the Schwinger field, $\hbar\omega_\gamma$ is the photon energy, $\mathbf{\hat k}$ is the direction of the photon wave vector, and the subscript $\perp$ ($\parallel$) represents the component perpendicular (parallel) to the motion of the electron or the photon.

If the radiation is treated as the emission of a series of incoherent photons, the differential photon emission rate can be obtained from $\mathrm{tr}\rho$ as
\begin{equation}
\frac{\mathrm{d}^2 N_\gamma}{\mathrm{d}\chi_\gamma\mathrm{d}t} = \frac{\mathrm{d}^2 [\mathrm{tr}\rho]}{\mathrm{d}\chi_\gamma\mathrm{d}t} = \frac{\alpha \mathscr{F}}{\sqrt{3} \pi\tau_c \gamma \chi_e},\label{dN}
\end{equation}
where 
\begin{equation}
\mathscr{F} = \left (2+3\chi_\gamma y \right ) K_{2/3}(2y) - \mathrm{Int}K(2y) - \frac{\chi_\gamma}{\chi_e} \Xi_\zeta K_{1/3}(2y). \nonumber
\end{equation}
The emission process is considered instantaneous, so the final polarization state after the emission of a Compton photon with $\chi_\gamma$ at time $t$ can be calculated from $\boldsymbol{\Xi} = \mathrm{tr}[\boldsymbol{\sigma}\rho(t, \chi_\gamma)]/\mathrm{tr}\rho(t, \chi_\gamma)$, and the polarization variation is
\begin{eqnarray}
\begin{aligned}
\Delta \Xi_\kappa &=  \frac{\Xi_\kappa}{\mathscr{F}} \left[ \frac{\chi_\gamma}{\chi_e} \Xi_\zeta K_{1/3}(2y) - 3\chi_\gamma y \mathrm{Int}K(2y) \right], \\
\Delta \Xi_\eta &= \frac{\Xi_\eta}{\mathscr{F}} \left[ \frac{\chi_\gamma}{\chi_e} \Xi_\zeta K_{1/3}(2y) - 3\chi_\gamma y K_{2/3}(2y) \right], \\
\Delta \Xi_\zeta &= \frac{1}{\mathscr{F}} \left[\left( \frac{\chi_\gamma}{\chi_e}\Xi_\zeta^2 - 3\chi_e y \right) K_{1/3}(2y) - 3\chi_\gamma y \Xi_\zeta K_{2/3}(2y) \right].
\end{aligned}\label{delta_xi}
\end{eqnarray}

To get an intuitive picture of the nonlinear Compton scattering, we image the quantum synchrotron emissivity $F=(2\chi_\gamma/3\chi_e^2)\mathscr{F}$ and the polarization variation $|\Delta \boldsymbol{\Xi}|$ after the emission of a photon in Fig.~\ref{fig:delta_xi} for different $\chi_e$, $\chi_\gamma$, and initial polarization $\boldsymbol{\Xi}$. The quantum synchrotron emissivity $F$ reflects the emission probability of different photon energy on a log scale of $\chi_\gamma$~\cite{erber1966high,duclous2010monte}. 
The polarization variation $|\Delta \boldsymbol{\Xi}|$ after the emission of a photon approximately increases with the power of $\chi_\gamma/\chi_e$. In the classical limit $\chi_e \ll 1$, the energy of the emitted photons is far less than that of the electron itself, so it is only possible to be significantly polarized or depolarized after a lot of photons are emitted.
However, for the quantum regime $\chi_e \gtrsim 1$, the emitted photons may have the same magnitude of energy as the electron so that the electron can be significantly polarized after merely a few photons are emitted.
Moreover, the emission probability distribution for $\chi_e \ll 1$ is almost the same for different polarizations, but when $\chi_e \gtrsim 1$, the emission probability for the polarization oriented in the opposite direction of $\boldsymbol{\zeta}$ is obviously larger than the one oriented in the same direction of $\boldsymbol{\zeta}$.

\subsection{Instantaneous no photon emission}
Through the density matrix, we obtain the final polarization vector after the emission of a Compton photon. 
However, when using the polarization vector to describe the spin state of the electron, we should be aware that the polarization vector varies under the influence of radiation even when there is instantaneously no photon emitted~\cite{cain2011user,li2020production}. 

The polarization vector of a single electron indicates the probability of different spin states in measurement. In other words, the polarization vector can be obtained from the measurement of a large number of electrons with the same state~\cite{landau1981quantum}. In order to calculate the polarization vector evolution of instantaneously no-photon-emitted electrons, we can investigate a collection of $N$ electrons with identical initial conditions. In a very short time interval, $\Delta t$, the probability of emitting multiple photons is negligible compared to emitting a single photon, so the total number of the electrons emitting a photon in $\Delta t$ can be calculated from Eq.~\eqref{dN} as
\begin{equation}
    N^{\mathrm{emit}} = N \left(\int_0^{\chi_e} \frac{\mathrm{d}^2 N_\gamma}{\mathrm{d}\chi_\gamma\mathrm{d}t} \mathrm{d}\chi_\gamma \right) \Delta t = N (\mathcal{F}_0 - \boldsymbol{\Xi}\cdot\boldsymbol{\zeta} \mathcal{F}_s)\Delta t,\label{N_emit}
\end{equation}
where
\begin{eqnarray}
\mathcal{F}_0 &=& \frac{\alpha}{\sqrt{3} \pi\tau_c \gamma \chi_e} \int_0^{\chi_e} \left [ \left (2+3\chi_\gamma y \right ) K_{2/3}(2y) - \mathrm{Int}K(2y) \right ] \mathrm{d}\chi_\gamma, \nonumber \\
\mathcal{F}_s &=& \frac{\alpha}{\sqrt{3} \pi\tau_c \gamma \chi_e} \int_0^{\chi_e} \frac{\chi_\gamma}{\chi_e} K_{1/3}(2y) \mathrm{d}\chi_\gamma. \nonumber
\end{eqnarray}
Apart from the emitting electrons, the population of the rest no-photon-emitted ones is
\begin{equation}
    N^{\mathrm{no}} = N - N^{\mathrm{emit}} = N \left[ 1 - \left(\mathcal{F}_0 - \boldsymbol{\Xi}\cdot\boldsymbol{\zeta} \mathcal{F}_s \right)\Delta t \right].\label{N_no}
\end{equation}

To obtain the final polarization vector of the no-photon-emitted electrons, we can arbitrarily take a quantization axis, $\mathbf{e}$, and then the spin is only allowed to be in two states, i.e., parallel ($+\mathbf{e}$) or antiparallel ($-\mathbf{e}$) to the axis. As the component of the polarization vector along a certain axis represents the mean spin value in measurement along this axis, the number of electrons in the two states $\pm \mathbf{e}$ can be solved as
\begin{equation}
    N_{\pm \mathbf{e}} = N \left(1 \pm \mathbf{e} \cdot \boldsymbol{\Xi} \right) / 2.
\end{equation}
Similar to Eqs.~\eqref{N_emit} and \eqref{N_no}, the number of no-photon-emitted electrons of the two spin states is expressed as
\begin{equation}
    N_{\pm \mathbf{e}}^{\mathrm{no}} = N_{\pm \mathbf{e}} \left[ 1 - (\mathcal{F}_0 \mp \mathbf{e}\cdot\boldsymbol{\zeta}\mathcal{F}_s)\Delta t \right].
\end{equation}
When $N$ is very large, the relative frequency is equal to the probability. We utilize the relative frequency to calculate the final mean spin value of the no-photon-emitted electrons along the axis $\mathbf{e}$, i.e.,
\begin{equation}
    \mathbf{e} \cdot \boldsymbol{\Xi}^{\mathrm{no}} = \frac{N_{+\mathbf{e}}^{\mathrm{no}}}{N^{\mathrm{no}}}\cdot(+1) + \frac{N_{-\mathbf{e}}^{\mathrm{no}}}{N^{\mathrm{no}}}\cdot(-1).
\end{equation}
Substituting $ N_{\pm \mathbf{e}}^{\mathrm{no}}$ and $N^{\mathrm{no}}$ into the above equation, we can obtain~\cite{cain2011user}
\begin{equation}
    \mathbf{e}\cdot\boldsymbol{\Xi}^{\mathrm{no}} = \frac{ \mathbf{e}\cdot[\boldsymbol{\Xi} \left( 1 - \mathcal{F}_0 \Delta t \right) + \boldsymbol{\zeta}\mathcal{F}_s \Delta t] }{1 - \left( \mathcal{F}_0 - \boldsymbol{\Xi} \cdot \boldsymbol{\zeta} \mathcal{F}_s \right) \Delta t}.\label{xi'_zeta}
\end{equation}
Noting that the axis $\mathbf{e}$ is arbitrary, we can take away $\mathbf{e}$ from both sides of the above equation and get the expression of the final polarization vector $\boldsymbol{\Xi}^{\mathrm{no}}$. Then, the polarization variation of the no-photon-emitted electrons in the time interval $\Delta t$ is
\begin{equation}
\Delta \boldsymbol{\Xi}^\mathrm{no} = \boldsymbol{\Xi}^\mathrm{no} - \boldsymbol{\Xi} =  \frac{\left[ \boldsymbol{\zeta} - (\boldsymbol{\Xi}\cdot\boldsymbol{\zeta})\boldsymbol{\Xi} \right] \mathcal{F}_s \Delta t}{1 - \left( \mathcal{F}_0 - \boldsymbol{\Xi}\cdot\boldsymbol{\zeta} \mathcal{F}_s \right) \Delta t}.
\end{equation}
Since the emission of photons is considered instantaneous, if we let $\Delta t \to 0$, the above variation will be applied to every electron. Thus, we obtain the time differential of the polarization vector variation due to instantaneous no photon emission as
\begin{equation}
\left(\frac{\mathrm{d}\boldsymbol{\Xi}}{\mathrm{d}t}\right)_{\mathrm{NP}} = \left(\frac{\Delta \boldsymbol{\Xi}^\mathrm{no}}{\Delta t}\right)_{\Delta t \to 0} = \left[ \boldsymbol{\zeta} - (\boldsymbol{\Xi}\cdot\boldsymbol{\zeta})\boldsymbol{\Xi} \right] \mathcal{F}_s. \label{d_xi_NP}
\end{equation}

\subsection{Classical spin precession}
The above discussion is all about the spin polarization evolution under the influence of radiation. On the other hand, the polarization vector is also dominated by the classical spin precession, which is described by the the Bargmann-Michel-Telegdi (BMT) equation in the theory of classical spin of motion, i.e.~\cite{bargmann1959precession,jackson1998classical},
\begin{eqnarray}
\left(\frac{\mathrm{d} \boldsymbol{\Xi}}{\mathrm{d}t}\right)_\mathrm{BMT} &=& \frac{e}{m_e c} \boldsymbol{\Xi} \times \bigg[ \left( a + \frac{1}{\gamma} \right) \mathbf{B} - \frac{a \gamma}{\gamma+1} \left( \boldsymbol{\beta} \cdot \mathbf{B} \right) \boldsymbol{\beta} \nonumber \\
& & - \left( a - \frac{1}{\gamma+1} \right) \boldsymbol{\beta} \times \mathbf{E} \bigg],\label{BMT}
\end{eqnarray}
where $a = \alpha/2\pi \approx 1.16\times 10^{-3}$ is the anomalous magnetic moment of the electron~\cite{schwinger1948quantum}. So considering the influence of radiation and precession simultaneously, the evolution of the polarization vector when the electron is not emitting a photon is
\begin{equation}
\frac{\mathrm{d}\boldsymbol{\Xi}}{\mathrm{d}t} = \left(\frac{\mathrm{d}\boldsymbol{\Xi}}{\mathrm{d}t}\right)_{\mathrm{NP}} + \left(\frac{\mathrm{d} \boldsymbol{\Xi}}{\mathrm{d}t}\right)_\mathrm{BMT}. \label{dxi/dt}
\end{equation}

\section{\label{sec:simulation-method}Simulation Method}
Based on the above theories of the polarization vector, a spin-considered MCPV model of high-energy photon emission is easy to implement by extending the normal Monte Carlo radiation method~\cite{duclous2010monte,ridgers2014modelling,gonoskov2015extended,lobet2016modeling}. Following the original Monte Carlo model, the final optical depth $\tau_f$ sampled from $\tau_f = -\log \xi_1$ is assigned to each electron, where $\xi_1$ is a uniform random number in $(0,1)$, and the initial optical depth $\tau$ is set to 0. The optical depth $\tau$ evolves following the integral
\begin{equation}
\frac{\mathrm{d}\tau}{\mathrm{d}t} = \int_0^{\chi_e}{ \frac{\mathrm{d}^2N_\gamma}{\mathrm{d}\chi_\gamma \mathrm{d}t} \mathrm{d}\chi_\gamma } = \mathcal{F}_0 - \boldsymbol{\Xi}\cdot\boldsymbol{\zeta} \mathcal{F}_s.
\end{equation}
Once the optical depth $\tau$ reaches the final optical depth $\tau_f$, the electron emits a photon, another final optical depth is assigned in the same way, and the optical depth is reset to $0$. The quantum parameter $\chi_\gamma$ of a single-photon emission is decided by another uniform random number $\xi_2$ in $(0,1)$ following the equation
$
\xi_2 = [{\int_0^{\chi_\gamma}{ ({\mathrm{d}^2N_\gamma}/{\mathrm{d}\chi_\gamma \mathrm{d}t})  \mathrm{d}\chi_\gamma }}]/[{\int_0^{\chi_e}{ ({\mathrm{d}^2N_\gamma}/{\mathrm{d}\chi_\gamma \mathrm{d}t})  \mathrm{d}\chi_\gamma }}].
$
In this way, the rate of the photon production in the simulation will be consistent with Eq.~\eqref{dN} when the sample is very large. As the Monte Carlo model is usually used to deal with the high-energy photon emission, the emitted photon can be considered as moving oriented in the direction of motion of the electron under this condition. Then the photon energy can be calculated by $\hbar\omega_\gamma = \gamma m_e c^2 \chi_\gamma/\chi_e$, and the electron momentum after the emission becomes $\mathbf{p}' = \mathbf{p}-(\hbar\omega_\gamma/c)\mathbf{\hat p}$ according to momentum conservation.

To deal with the evolution of the spin polarization, we assign each electron particle an individual polarization vector. The polarization vector evolves following Eq.~\eqref{dxi/dt}. Besides, once the electron emits a photon (i.e., $\tau$ reaches $\tau_f$), the polarization vector has a step variation according to Eq.~\eqref{delta_xi}. In this way, the electron contains the complete information on the spin polarization at any moment under the influence of radiation and precession simultaneously.

In the classical regime $\chi_e \ll 1$, the radiative polarization phenomenon, as well as the radiation reaction, is only significant for the cumulative effect of incoherent photon emission. So under this condition, we can integrate the polarization variation for different emitted photon energies, and then the continuous expression of the polarization evolution under the influence of radiation can be written as
\begin{equation}
\left(\frac{\mathrm{d}\boldsymbol{\Xi}}{\mathrm{d}t}\right)_{\mathrm{R}} = \int_0^{\chi_e} \Delta \boldsymbol{\Xi} \frac{\mathrm{d}^2N_\gamma}{\mathrm{d}\chi_\gamma \mathrm{d}t} \mathrm{d}\chi_\gamma + \left(\frac{\mathrm{d}\boldsymbol{\Xi}}{\mathrm{d}t}\right)_{\mathrm{NP}}, \label{xi'}
\end{equation}
where $\Delta \boldsymbol{\Xi}$ is the polarization variation shown in Eq.~\eqref{delta_xi} and ${\mathrm{d}^2N_\gamma}/({\mathrm{d}\chi_\gamma \mathrm{d}t})$ is the differential photon emission rate shown in Eq.~\eqref{dN}. We can obtain the final continuous expression of the polarization evolution by substituting these equations:
\begin{equation}
\begin{split}
\left(\frac{\mathrm{d}\Xi_\zeta}{\mathrm{d}t}\right)_{\mathrm{R}} &= - \Xi_\zeta \mathcal{F}_2 - \mathcal{F}_1, \\
\left(\frac{\mathrm{d}\Xi_\eta}{\mathrm{d}t}\right)_{\mathrm{R}} &= - \Xi_\eta \mathcal{F}_2, \\
\left(\frac{\mathrm{d}\Xi_\kappa}{\mathrm{d}t}\right)_{\mathrm{R}} &= - \Xi_\kappa \mathcal{F}_3,
\end{split}\label{eq:continuous}
\end{equation}
where
\begin{eqnarray}
\mathcal{F}_1 &=& \frac{\alpha}{\sqrt{3} \pi\tau_c \gamma \chi_e} \int_0^{\chi_e} 3 \chi_\gamma y K_{1/3}(2y) \mathrm{d}\chi_\gamma, \nonumber \\
\mathcal{F}_2 &=& \frac{\alpha}{\sqrt{3} \pi\tau_c \gamma \chi_e} \int_0^{\chi_e} 3 \chi_\gamma y K_{2/3}(2y) \mathrm{d}\chi_\gamma, \nonumber \\
\mathcal{F}_3 &=& \frac{\alpha}{\sqrt{3} \pi\tau_c \gamma \chi_e} \int_0^{\chi_e} 3 \chi_\gamma y \mathrm{Int}K(2y) \mathrm{d}\chi_\gamma. \nonumber
\end{eqnarray}
This continuous expression completely agrees with the corresponding formula in Refs.~\cite{geng2019generalizing,guo2020stochasticity}, which is also evidence for the correctness of our MCPV model.

\section{\label{sec:results-discussions}Results and Discussions}
\subsection{\label{sec:sokolov-ternov}Sokolov-Ternov effect}
\begin{figure}
\includegraphics[keepaspectratio=true,width=86mm]{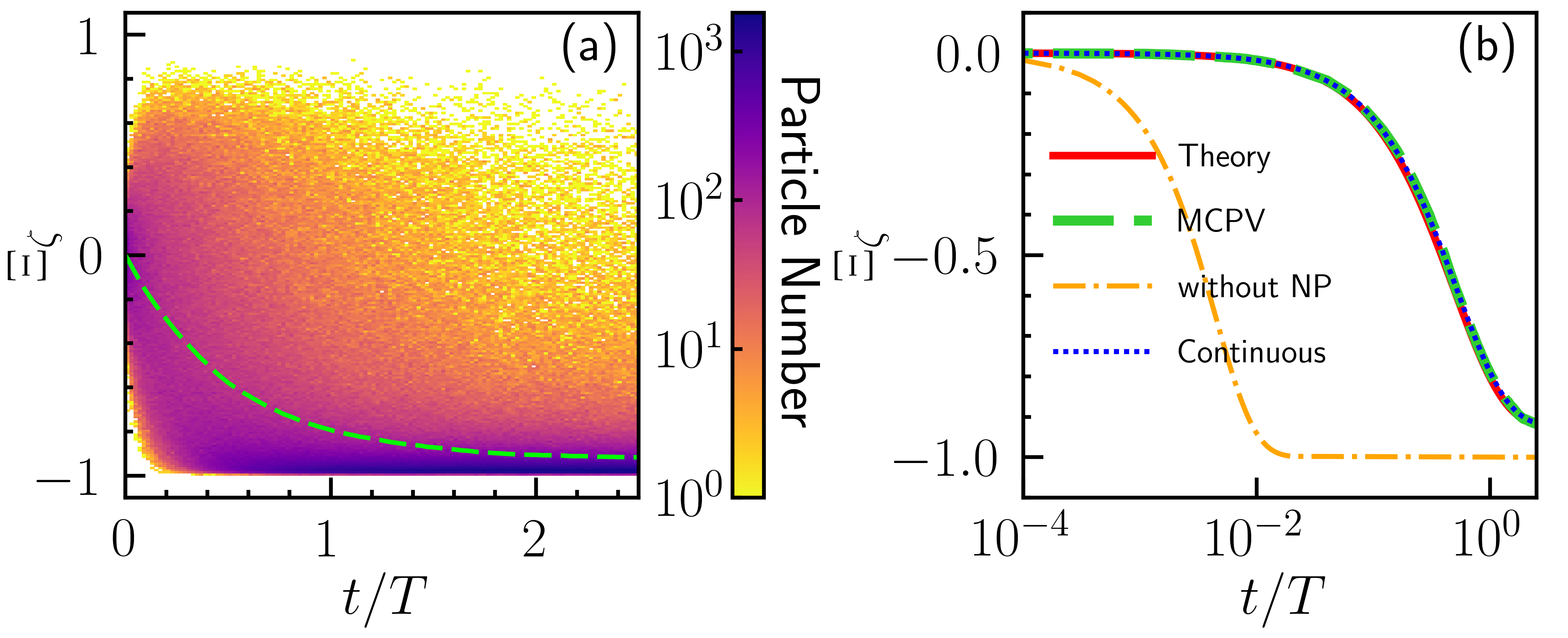}
\caption{\label{fig:Sokolov-Ternov} The simulation results of the Sokolov-Ternov effect. (a) The polarization distribution vs the simulation time for the MCPV model. The green dashed line is the average polarization. (b) The comparison of the polarization evolution of different models. The red solid line is the theoretical curve calculated from Eq.~\eqref{st_effect}. The blue dotted line is calculated from the continuous expression, i.e., Eq.~\eqref{eq:continuous}. The green dashed line is the average polarization of the MCPV model. The yellow dash-dotted line is the same as the green dashed line except that $(\mathrm{d}\boldsymbol{\Xi}/\mathrm{d}t)_{\mathrm{NP}}$ is not included in the simulation.}
\end{figure}
An initially unpolarized beam of high-energy electrons circulating in a storage ring will finally have a high-degree polarization oriented in the direction opposite to the magnetic field after a sufficiently long time, which is often called the Sokolov–Ternov effect. In this case, the polarization evolves following~\cite{sokolov1986radiation}
\begin{equation}
\Xi_\zeta = A (1 - e^{-t/\tau_r}), \label{st_effect}
\end{equation}
where $A$ is the limiting degree of polarization and $\tau_r$ is the relaxation time. For $\chi_e \ll 1$, there are $A = - 8\sqrt{3}/15 \approx -0.924$ and $\tau_r = (|A| \tau_c / \alpha \gamma^2) (B_{\mathrm{crit}}/B)^3$, where $B_{\mathrm{crit}} = 4.41 \times 10^9 \mathrm{T}$ is the Schwinger field.

Therefore, to check the correctness of our model, we first simulate 10000 particles with the same initial state using a Monte Carlo code to show the Sokolov–Ternov effect. The initial state of the electron is set to $\boldsymbol{\Xi} = 0$, $\gamma = 2 \times 10^6$, and moving perpendicular to the magnetic field. The magnetic field is a uniform and stationary field with $B = 10 \mathrm{T}$.
Then we can calculate that $\chi_e = 4.53 \times 10^{-3}$, the relaxation time $\tau_r = 3.50 \times 10^{-6} \mathrm{s}$, the circulating period $T = 7.14 \times 10^{-6} \mathrm{s}$, and the average emission time $\bar{\tau} \approx \mathcal{F}_0^{-1} = 5.39 \times 10^{-11} \mathrm{s}$.
So the time step is selected as $\Delta t = 1 \times 10^{-12} \mathrm{s} \ll \bar{\tau}$. To compare with the theoretical formula conveniently, the radiation reaction recoil is not switched on in the simulation.

The simulation result is shown in Fig.~\ref{fig:Sokolov-Ternov}. The polarization evolution of a single particle is highly dominated by the emission moment and the energy of its emitted photons. Therefore, the polarization of each particle is different at a certain moment, although they have the same initial state, which is shown in Fig.~\ref{fig:Sokolov-Ternov}(a). However, the average polarization of all the particles (the green dashed line) is in good agreement with the theoretical curve calculated from Eq.~\eqref{st_effect} (the red solid line) as shown in Fig.~\ref{fig:Sokolov-Ternov}(b). Besides, the evolution calculated from the continuous expression, Eq.~\eqref{eq:continuous} (the blue dotted line), is also consistent with the theoretical curve. In fact, Eq.~\eqref{eq:continuous} is completely equivalent to Eq.~\eqref{st_effect} with $A = - \mathcal{F}_1 / \mathcal{F}_2$ and $\tau_r = 1/\mathcal{F}_2$ under this condition.

We also run an identical simulation except that $(\mathrm{d}\boldsymbol{\Xi}/\mathrm{d}t)_{\mathrm{NP}}$ is not included in the code, and the result of the average polarization of 10000 particles is shown as the yellow dash-dotted line in Fig.~\ref{fig:Sokolov-Ternov}(b). It evolves with the characteristic time $\tau \approx 10^{-8}\mathrm{s} \ll \tau_r$ and the limiting polarization $\Xi_\zeta = -1 \ne -0.924$, which is completely different from the theory. It indicates that it is important to include $(\mathrm{d}\boldsymbol{\Xi}/\mathrm{d}t)_{\mathrm{NP}}$ in the simulation code to obtain the right results.

\subsection{Longitudinal polarization in magnetic field}
\begin{figure}
\includegraphics[keepaspectratio=true,width=86mm]{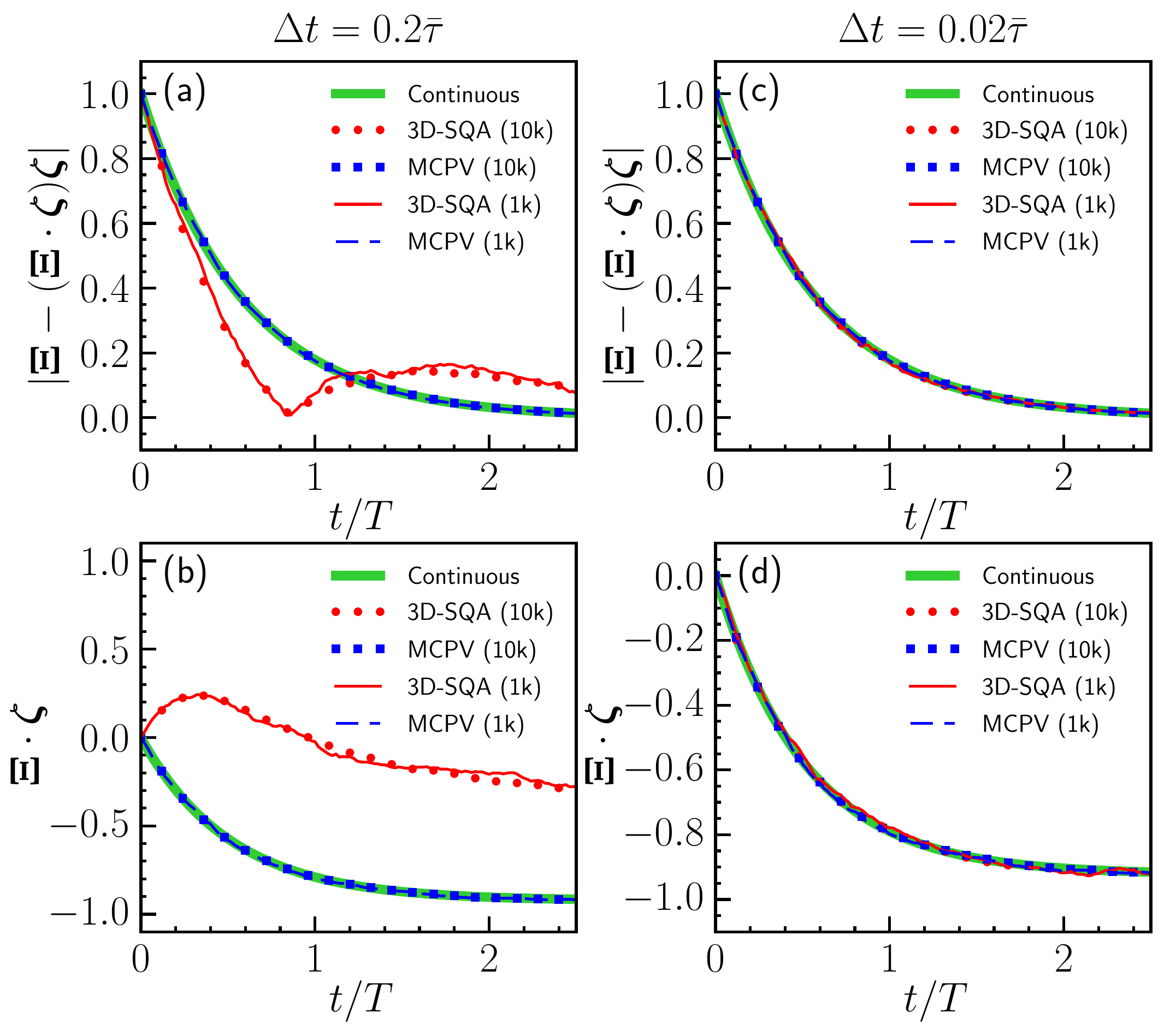}
\caption{\label{fig:LPCSMF} The polarization evolution of the electron with initially longitudinal polarization $\boldsymbol{\Xi} = \boldsymbol{\hat{\beta}}$ for time steps [panels (a) and (b)] $\Delta t = 0.2 \bar{\tau}$ and [panels (c) and (d)] $\Delta t = 0.02 \bar{\tau}$. The green thick lines, the red circular-marker lines, the blue square-marker lines, the red solid lines, and the blue dashed lines are the results of the continuous model, the average results of 10000 particles of the 3D-SQA model and the MCPV model, and the average results of 1000 particles of the 3D-SQA model and the MCPV model, respectively.}
\end{figure}
One of the major advantages of the MCPV model is that it can simulate the polarization evolution of an arbitrary axis simultaneously, so next we run a simulation where all the parameters are the same as the simulation in Sec.~\ref{sec:sokolov-ternov} except that the electron has initially longitudinal polarization, i.e., $\boldsymbol{\Xi} = \boldsymbol{\hat{\beta}}$. 
To demonstrate the accuracy and efficiency of the MCPV model, we also run the same simulation with the continuous model and the 3D-SQA model. 
In the continuous model, the polarization vector evolves following ${\mathrm{d}\boldsymbol{\Xi}}/{\mathrm{d}t} =\left({\mathrm{d}\boldsymbol{\Xi}}/{\mathrm{d}t}\right)_{\mathrm{R}} + \left({\mathrm{d} \boldsymbol{\Xi}}/{\mathrm{d}t}\right)_\mathrm{BMT}$. 
In the 3D-SQA model, the treatment of the electron spin is carried out according to the method in Ref.~\cite{li2020production}; i.e., the electron spin jumps to the state either parallel or antiparallel to the axis of the mean spin vector in each time step for both emitting and no-emitting electrons.

The simulation results are shown in Fig.~\ref{fig:LPCSMF}. Under this condition, $\chi_e = 4.53 \times 10^{-3} \ll 1$, so the results should all be consistent with the continuous model. The electron is depolarized in the plane perpendicular to the magnetic field and simultaneously polarized in the direction opposite to the magnetic field.
The results of the MCPV model, whether $\Delta t = 0.2\bar{\tau}$ or $\Delta t = 0.02\bar{\tau}$, 1000 particles averaged or 10000 particles averaged, all agree well with each other and the continuous model. However, the results of $\Delta t = 0.2\bar{\tau}$ of the 3D-SQA model are completely inconsistent with the results of $\Delta t = 0.02\bar{\tau}$ and other models, which indicates that the 3D-SQA model needs a stricter time step condition to obtain the correct results.
For $\Delta t = 0.02\bar{\tau}$, although the consistent results can be obtained through the 3D-SQA model, there are still obvious differences in accuracy between 1000 and 10000 particles averaged results of the 3D-SQA model, while there are no significant differences for the MCPV model.
Therefore, it indicates that the MCPV model requires a more relaxed condition of time step and less statistic particles to obtain the correct and accurate results than the 3D-SQA model, which demonstrates the computational efficiency advantage of the MCPV model.

\subsection{Longitudinal polarization in laser field}
\begin{figure}
\includegraphics[keepaspectratio=true,width=86mm]{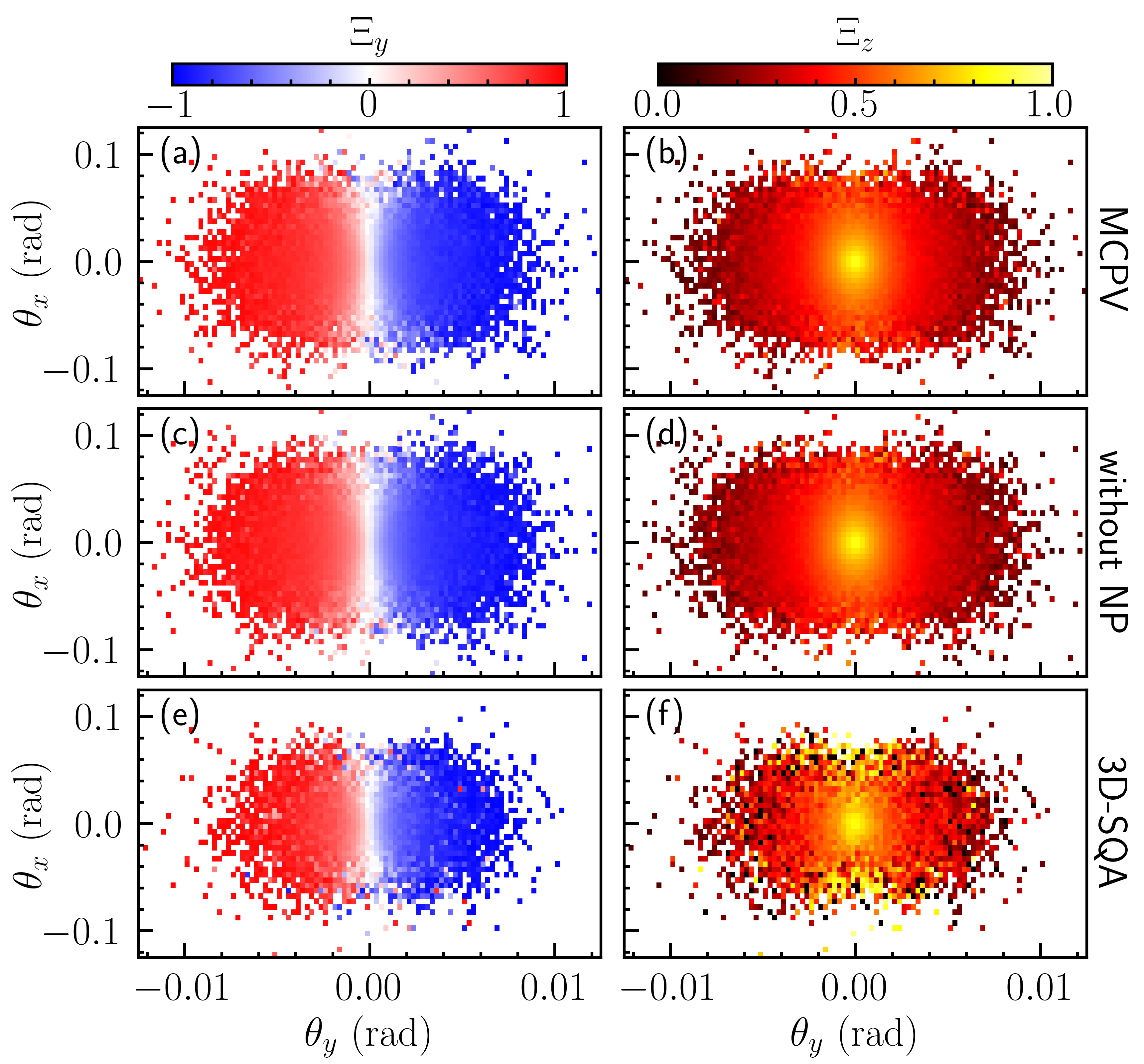}
\caption{\label{fig:laser} The distribution of the average polarization at different deflection angles for [panels (a) and (b)] the MCPV model, [panels (c) and (d)] the polarization vector model not including $(\mathrm{d}\boldsymbol{\Xi}/\mathrm{d}t)_{\mathrm{NP}}$, and [panels (e) and (f)] the 3D-SQA model.}
\end{figure}
\begin{figure}
\includegraphics[keepaspectratio=true,width=86mm]{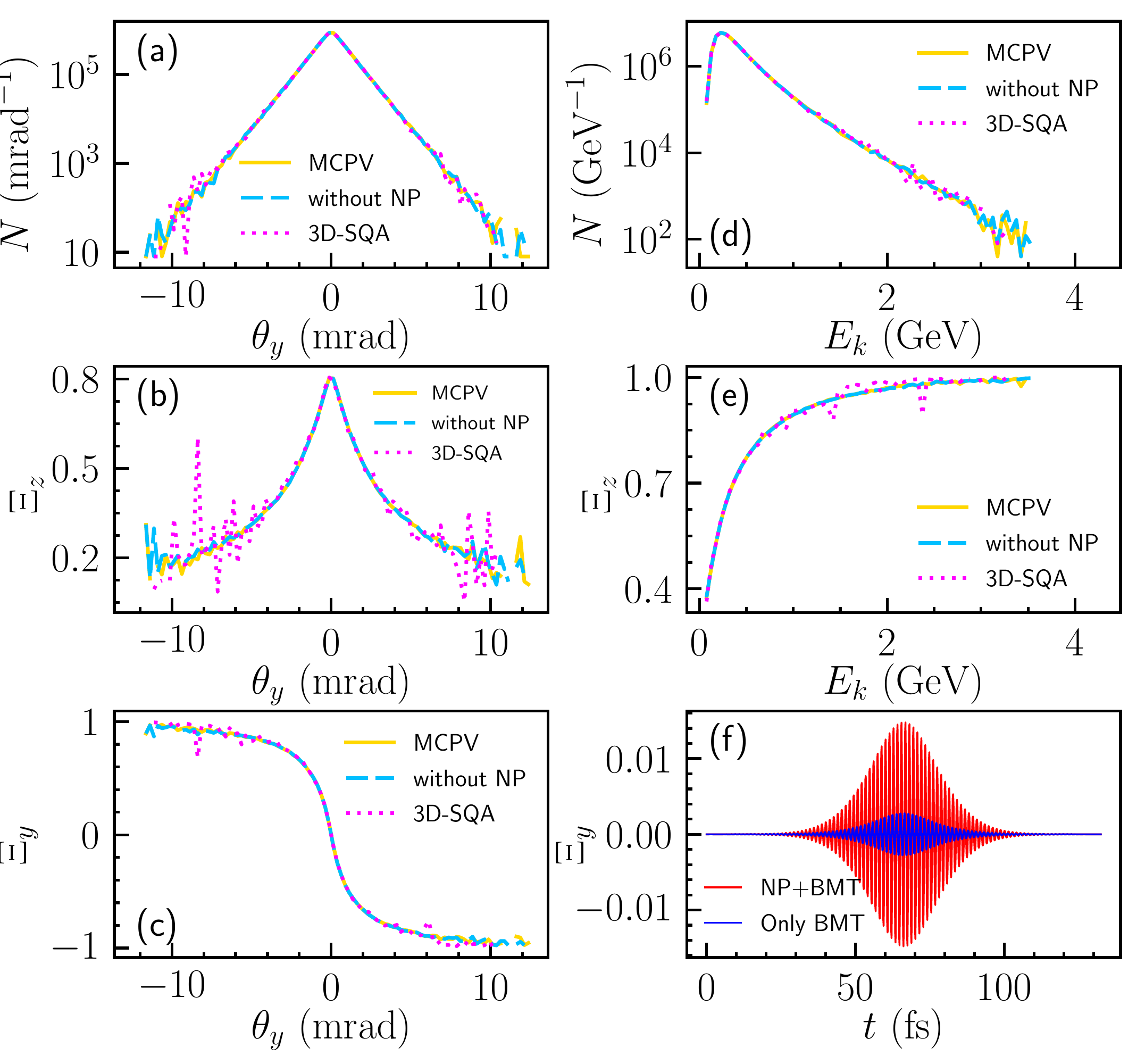}
\caption{\label{fig:laser_plot} The distribution of (a) the particle number, (b) the average polarization along the $z$ axis, and (c) the average polarization along the $y$ axis at different deflection angles $\theta_y$. The distribution of (d) the particle number and (e) the average polarization along the $z$ axis for different kinetic energies after collision. The yellow solid lines, the blue dashed lines, and the pink dotted lines are the results of the MCPV model, the polarization vector model not including $(\mathrm{d}\boldsymbol{\Xi}/\mathrm{d}t)_{\mathrm{NP}}$, and the 3D-SQA model, respectively. (f) The polarization evolution induced by $(\mathrm{d}\boldsymbol{\Xi}/\mathrm{d}t)_{\mathrm{NP}}$ and precession simultaneously (the red line) and only precession (the blue line).}
\end{figure}
To comprehensively investigate the applicability of the MCPV model, we also run a simulation in a laser field. According to Ref.~\cite{li2019ultrarelativistic}, an ultraintense elliptically polarized laser pulse can split a relativistic electron beam along the propagation direction into two transversely polarized parts. The electron particle is initialized to propagate along the $-z$ direction with the kinetic energy $E_k = 5 \mathrm{GeV}$ and the longitudinal polarization $\boldsymbol{\Xi} = \mathbf{\hat{z}}$ at the beginning of the simulation. A Gaussian laser pulse whose center axis coincides with the $z$ axis propagates along the $+z$ direction and head-on collides with the electron at the focus of the laser pulse. The laser peak intensity $I_0 = 5.34 \times 10^{21} \mathrm{W/cm^2}$ $(a_0 = 50)$, the wavelength $\lambda = 800 \mathrm{nm}$, the pulse duration $\tau = 30 \mathrm{fs}$, the waist radius $w_0 = 5.0 \mathrm{\mu m}$, and the ellipticity $\epsilon = |E_y|/|E_x| = 0.05$. In this case, the max electron quantum parameter is $\chi_e = 1.5$. We also use the 3D-SQA model and the polarization vector model not including $(\mathrm{d}\boldsymbol{\Xi}/\mathrm{d}t)_{\mathrm{NP}}$ for comparison.
The time step is set as $\Delta t = 2.67 \times 10^{-3} \mathrm{fs}$ and the number of statistic particles is $1 \times 10^6$ for each model.

The simulation results of all the models are consistent with each other and agree with the conclusions in the literature, as shown in Figs.~\ref{fig:laser} and \ref{fig:laser_plot}. After colliding with the laser pulse, the electron particles propagate along different directions due to the stochasticity in the quantum radiation recoil  regime. It also implies that if we use the continuous model for this case, all particles will always be in the same state since there is no stochastic process, so the continuous model is not suitable for this case. The average spin polarization varies with the deflection angle after collision. For the polarization along the $y$ axis, i.e., $\Xi_y$, the degree of polarization increases with the increase of the deflection angle in the $\pm y$ directions, i.e., $|\theta_y|$. The electrons have a degree of polarization in the $-y$ direction when $\theta_y > 0$ and have a degree of polarization in the $+y$ direction when $\theta_y < 0$. For the initial polarization direction $z$, the degree of depolarization increases with the increase of the deflection angle, especially in the $\pm y$ direction, and the electron with smaller energy after collision has a higher degree of depolarization.

What is completely different from the results in Sec.~\ref{sec:sokolov-ternov} is that the same result is obtained whether $(\mathrm{d}\boldsymbol{\Xi}/\mathrm{d}t)_{\mathrm{NP}}$ is considered or not. In other words, in this case, the right result can be obtained even if $(\mathrm{d}\boldsymbol{\Xi}/\mathrm{d}t)_{\mathrm{NP}}$ is not included in the simulation code. Since the direction of the magnetic field of the laser changes periodically, the polarization induced by $(\mathrm{d}\boldsymbol{\Xi}/\mathrm{d}t)_{\mathrm{NP}}$ also oscillates periodically as shown in Fig.~\ref{fig:laser_plot}(f). Therefore, the polarization variation caused by $(\mathrm{d}\boldsymbol{\Xi}/\mathrm{d}t)_{\mathrm{NP}}$ does not accumulate in a certain direction, and its amplitude is usually much less than the polarization variation when the electron emits a high-energy photon. Based on the above reasons, for the electromagnetic field whose direction changes fast and periodically as the laser field, it is feasible to neglect the influence of $(\mathrm{d}\boldsymbol{\Xi}/\mathrm{d}t)_{\mathrm{NP}}$. However, for a constant field like the one in Sec.~\ref{sec:sokolov-ternov}, the polarization variation caused by $(\mathrm{d}\boldsymbol{\Xi}/\mathrm{d}t)_{\mathrm{NP}}$ accumulates in a certain direction and the inaccurate result will be obtained if $(\mathrm{d}\boldsymbol{\Xi}/\mathrm{d}t)_{\mathrm{NP}}$ is not included in the simulation code. The electromagnetic field in the plasma environment is very complicated, so the model not including the polarization effect due to instantaneous no photon emission may lead to incorrect results when applied in plasma simulation such as the PIC code. In contrast, our MCPV model can obtain accurate results in such an electromagnetic environment and has high computational efficiency, and thus it can be well integrated into the PIC simulation to deal with the spin polarization.

Although both the MCPV model and the 3D-SQA model can give the correct results, they are completely different from the perspective of the quantum physical picture. There is a quantum uncertainty relation between the spin in orthogonal directions. In other words, the spins in orthogonal directions cannot be measured simultaneously since they are not commutative. Therefore, the spin is not in a definite state but only a probability distribution (i.e., the polarization vector) before the actual measurement. The MCPV model follows this physical picture and treats the effect of radiation on spin as the change in probability. However, the process in the 3D-SQA model that lets the spin fall on a certain state along the SQA in each time step is against the physical picture. It just uses the Monte Carlo method to make the results of spin approach the actual value by the calculation of a lot of particles. That is why the 3D-SQA model has larger fluctuations in the area with fewer particles as shown in Figs.~\ref{fig:laser} and \ref{fig:laser_plot}, which is also evidence that the MCPV model requires less particles than the 3D-SQA model to obtain equally accurate results.

\section{\label{sec:conclusion}Conclusion}
In conclusion, we propose a self-consistent polarization vector model to theoretically describe the spin of relativistic electrons in an intense electromagnetic field, and we develop a Monte Carlo method to simulate the electron spin under the influence of radiation and precession simultaneously. The model extends the applicability of the polarization vector model to wider electromagnetic environments by including the variation of radiative polarization due to the instantaneous no photon emission in the model. In the MCPV model, the electron contains the complete information on the spin polarization at any moment under the influence of radiation and precession simultaneously. The Sokolov-Ternov effect is successfully reproduced using our MCPV model, and the MCPV model shows good agreement with the continuous model in the classical regime. Compared with the 3D-SQA model, the MCPV model has higher computational efficiency and gets more accurate results under the same computational conditions. Based on the wide applicability and high computational efficiency, the MCPV model is a superior choice for PIC simulation to deal with the spin polarization.

\begin{acknowledgments}
The work has been supported by the NSFC (Grants No. 11921006 and No. 11535001) and the National Grand Instrument Project (Grant No. 2019YFF01014400). J.Y. has been supported by the project of Science and Technology on Plasma Physics Laboratory (Grant No. 6142A04190111) and the Natural Science Foundation of Hunan Province (Grant No. 2020JJ5031). Simulations were supported by the High-Performance Computing Platform of Peking University.
\end{acknowledgments}


\bibliography{apssamp}

\end{document}